\documentclass[12pt]{article}
\RequirePackage[OT1]{fontenc}
\RequirePackage{amsthm,amsmath}
\usepackage{accents}
\usepackage{breqn}
\usepackage{soul}
\usepackage[numbers,sort&compress]{natbib}
\RequirePackage[citecolor=blue,urlcolor=blue]{hyperref}
\usepackage{hyperref}
\usepackage{amsfonts}
\usepackage{amssymb}
\usepackage{graphics}
\usepackage{setspace}
\usepackage{physics}

\numberwithin{equation}{section}
\newcommand{\brho}{{\boldsymbol \rho}}

\newcommand{\bPsi}{{\boldsymbol \Psi}}

\def\b1{{\mathbf 1}}
\def\b0{{\mathbf 0}}

\def\bA{{\boldsymbol A}}

\def\bu{{\boldsymbol u}}

\def\bV{{\boldsymbol V}}

\def\bq{{\boldsymbol q}}

\def\bH{{\boldsymbol H}}

\def\bU{{\boldsymbol U}}

\def\cA{{\mathcal A}}

\def\cbU{{\mathcal {\boldsymbol{U}}}}
\def\cbA{{\mathcal {\boldsymbol{A}}}}
\def\cbV{{\mathcal {\boldsymbol{V}}}}

\def\cH{{\mathcal H}}
\def\bcH{{\boldsymbol \cH}}

\def\cS{{\mathcal S}}

\def\cG{{\mathcal G}}
\def\bbR{{\mathbb R}}

\def\bbI{{\mathbb I}}

\def\bbH{{\mathbb H}}
\def\bbC{{\mathbb C}}

\def\fU{{\mathfrak U}}

\begin{document}

\title{Quantum systems in Markovian environments}
\author{Henryk Gzyl\\
Centro de Finanzas IESA, Caracas, Venezuela.\\
 henryk.gzyl@iesa.edu.ve}
\date{}
 \maketitle

\setlength{\textwidth}{4in}

\vskip 1 truecm
\baselineskip=1.5 \baselineskip \setlength{\textwidth}{6in}

\begin{abstract}
In this work, we develop a mathematical framework to model a quantum system whose Hamiltonian may depend on the state of changing environment, that evolves according to a Markovian process. When the environment changes its state, the quantum system may suffer a shock that produces an instantaneous transition among its states. The model that we propose can be readily adapted to more general settings.\\
To avoid collateral analytical issues, we consider the case of quantum systems with finite dimensional state space, in which case the observables are described by Hermitian matrices. We show how to average over the environment to predict the expected values of observables.
\end{abstract}

\noindent \textbf{Keywords}: Quantum systems in random environments, Random evolutions, Systems subject to random pulses. Instantaneous perturbations. \\


\section{Introduction and Preliminaries} 
This paper is inspired by the work by Clauser and Blume, \cite{CB}. They consider a finite-dimensional system, whose time evolution is described by a Hamiltonian $H_0,$ that is subject to random pulses occurring at the jump times of a Poisson process of intensity $\lambda,$ at which a perturbation that changes the state of the systems instantaneously acts, after which the system again evolves according to $H_0.$ Momentarily we will explain the meaning of an instantaneous change of state.

Let us begin by saying that this setup admits several possible extensions, one of which is considered here, namely, that consisting of a finite-dimensional quantum system, that evolves in a time-varying environment, whose state is governed by a Markov process $\xi(t)$ having a finite state space $\cS=\{1,...,K\},$ and time-homogeneous transition probability $P_t(i,j).$ When the environment is in the $i$-th state, the quantum system evolves according to a Hamiltonian $H_i.$

Before further explanations, let us mention that the models of dynamical systems in random (Markovian) environments exist since quite a while. Probably, the origin of such models goes back to \cite{Gol}, who was interested in fluid dynamics. But that model can also be thought of as a model for physical Brownian motion that does not violate relativistic causality. In this regard see as well \cite{CV}. The subject eventually became a class of random models in itself. See \cite{Pin}, \cite{Swi}, and \cite{Kol} for details, history, references, and applications of dynamical systems in Markovian (or more general) environments. The Markov property implies that the jump times between states are exponentially distributed. There exists a larger class models, for which this is not necessarily true. It comprises those considered in \cite{Swi}, and known as \textit{random evolutions or piecewise deterministic dynamical systems}. As in the model that we consider here, those models would consist of a collection of quantum systems, modeled by a collection of possibly different Hamiltonians, but now the random times between switchings are not necessarily exponentially distributed. That is, the environment does not necessarily evolve according to a time-homogeneous Markov chain. See \cite{Da} for the basic mathematics and interesting applications. The models that we consider here fit into the larger class of stochastic modeling of open systems, but do not seem to have been analyzed before. See \cite{vK} and \cite{LW} for two comprehensive treatises of open systems, and \cite{LR} for the fluctuation-dissipation relations and, \cite{Cz}, \cite{GSKC} for two recent works on the use of stochastic methods to describe quantum open systems. For a probabilistic approach of particles in random media, consider \cite{GGN} and \cite{BGN}.

The paper is organized as follows. In the remainder of this section, to establish the basic notations, we recall some basic facts about Markov process on a finite state space. Then we explain how a quantum system evolves and what it is that we mean by instantaneous change of state. To make predictions about the system, we have to take care of two issues. First, given an initial state $\psi(0),$ we have to know what the future state and what the mean value of the observables of interest are, for any given history of the random environment. Then, we have to average both over all possible histories of the environment, that is over all random paths of the environment process $\xi(t).$ All of that is carried out in section 2.

In Section 3 we consider some examples. The first one, consists of a two-dimensional quantum system evolving in a two state environment. The second consists of a two-dimensional quantum system in a one state environment, in which the emphasis is on the instantaneous perturbation. In both cases we carry out the computations analytically almost to the end.

 \subsection{Specifying the environment}
A Markov chain $\xi(t)$ on $\cS=\{1,...,K\}$ is specified by its transition semigroup $P_t(i,j)=P(\xi(t)=j|\xi(0)=i).$
The average over random paths is defined on functions defined on collections of random sequences $(\xi(t_1),\xi(t_2),...,\xi(t_n))$ for $0\leq t_1\leq ... \leq t_n$ and any $f:\cS^{n}\to\bbR$ by:
 \begin{equation}\label{MP1}
\begin{aligned}
&E^{i}[f(\xi(t_1),\xi(t_2),...\xi(t_n))]\\ 
&= \sum_{j_n}\sum_{j_{n-1}}...\sum_{j_1}f(j_1,j_2,...,j_n)P_{t_1}(i,j_1)P_{t_2-t_1}(j_1,j_2)...P_{t_n-t_{n-1}}(j_{n-1},j_n).
\end{aligned}
\end{equation}
From that, plus successive approximations, one builds the probability on the space of all right continuous paths on $\cS.$ It is rather convenient to think of the initial state as  a random variable, namely, $\xi(0);$  
then regard $E^{\xi(0)}[F]=E[F|\xi(0)]$  as a random variable that acquires a value when the state of the environment is observed. Therefore, $E^{i}[F]$ means $E[F|\xi(0)=i],$  which describes the average of the random variable $F$ over all trajectories starting from $i\in\cS.$

The Markov property is stated as follows. Let $f:\cS^{n}\to\bbR,$ and $0\leq s_1\leq s_2\leq...\leq s_n\leq s\leq t_1\leq,,,,\leq t_n.$ Then: 
\begin{equation}\label{MP2}
\begin{aligned}
&E^{i}[f(\xi(t_1),\xi(t_2),...\xi(t_n))|\xi(s_1),...,\xi(s_n),\xi(s)]\\
& = E^{\xi(s)}[f(\xi(t_1-s),\xi(t_2-s),...\xi(t_n-s))]
\end{aligned}
\end{equation}

This is often stated as: To predict the future, it suffices to know the present. Implicitly, we are equating predicting to the computation of an expected (and more generally, of a conditional expected) value. 

Two important constructs that come up below are the following. First, the (random) time of the first jump is defined as
\begin{equation}\label{JT1}
T = \min\{t>0|\xi(t)\not= \xi(0)\}.
\end{equation}
The successive jump times are defined as follows. Set $T_0=0$ and $T_1=T,$ and put:
\begin{equation}\label{JT2}
T_{n+1} = \min\{t>T_n|\xi(t)\not= \xi(T_n)\},\;\;\;n\geq 0.
\end{equation}

It follows from the Markov property that $P^{i}(T>t)=e^{-\lambda_i t}$ for some $\lambda_i>0.$ By the way, this implies that $P^{i}(0<T<\infty)=1$ at any state. It can also be proved that the times between jumps,  $T_{n+1}-T_n,$ are statistically independent random variables. The \textit{first jump distribution} is defined by
\begin{equation}\label{JT3}
Q(i,j) = P^{i}(\xi(T)=j) = P(\xi(T)=j|\xi(0)=i).
\end{equation}

This implies that $\sum_{j\not=i}Q(i,j)=1.$ With these notations, it is convenient to specify the infinitesimal generator of $P_t$ as follows. Let $h:\cS\to\bbR$ be given and put $v(i,t)=E^{i}[h(\xi(t)].$ Then $v(i,t)$ is the probabilistic representation of the solution to
\begin{equation}\label{IG}
\frac{d v(i,t)}{dt} \equiv \cG v(i,t) = \lambda_i \sum_{j\not= i}Q(i,j)v(j,t) -\lambda_iv(i,t)= \sum_{j=1}^nq(i,j)v(j,t),\;\;\;v(i,0)=h(i).
\end{equation}
Here we wrote $q(i,j)=\lambda_iQ(i,j)$ for $i\not= j$ and $q(i,i)=-\lambda_i=-\sum_{j\not= i}q(i,j).$  The matrix $q(i,j)$ is also called the transition rate matrix. We just made explicit its intrinsic probabilistic structure. 

To close this section, we bring forth another important tool in probabilistic analysis: the {\it first jump time decomposition}. If $\cA$ denotes an event related to the environment, and $F$ is some random variable, we use
$$E^{i}[F;\cA] = \int_{cA} FdP^{i}$$
to denote its average over all the trajectories of the environment compatible with $\cA.$ Let now $\cA=\{T>t\}$ denote the event ``the environment has not changed its state up to time $t$'', and let $\{0\leq T \leq t\}$ denote its complement. Then,
$$v(i,t) = E^{i}[h(\xi(t));T>t] + E^{i}[h(\xi(t));0<T\leq t] .$$
Since the environment is in state $\xi(0)=i$ up to the first jump time, we have
$$E^{i}[h(\xi(t);T>t]  = h(i)P^{i}\big(T>t\big) = h(i)e^{-\lambda_i t}.$$
Invoking the Markov property at the time of the first jump, and the fact that the jump time is exponentially distributed, we have
$$E^{i}[h(\xi(t));0<T\leq t] = \lambda_i\sum_{j\not=i}\int_0^t e^{-\lambda_i s}Q(i,j)E^{j}[h(\xi(t-s)]ds =  \lambda_i\sum_{j\not=i}\int_0^t e^{-\lambda_i s}Q(i,j)v(j,t-s)ds.$$
Replacing $t-s$ by $s$ and putting the two identities together we have the first-time decomposition of $v(i,t):$

\begin{equation}\label{FTD1}
v(i,t) = h(i)e^{-\lambda_i t} + \lambda_i\sum_{j\not=i}\int_0^t e^{-\lambda_i(t-s)}Q(i,j)v(j,s)ds.
\end{equation}
If one differentiates \eqref{FTD1} with respect to time, we obtain \eqref{IG}.

\subsection{The quantum dynamics}
Let $\bbH=\{\psi\in\bbC^N| \langle\psi,\psi\rangle=1\}$ denote the states of the (finite-dimensional) quantum system. We use the standard notation $\langle\phi,\psi\rangle=\sum_{j}\bar{\phi}_j\psi_i$ to denote the scalar product in $\bbC^N.$ To describe the dependence on the environment we find it convenient to think of the states of the joint system (environment plus quantum system) as a map $\bPsi:[0,\infty)\times\cS \to \bbH_K,$ where $\bbH_K=\oplus\bbH$ is the direct sum of $\bbH$ $K$-times. For $1\leq i\leq K,$ $\psi(t, i)$ is the state of the quantum system at time $t$ when the environment is in state $i\in\cS.$ To think of the state vector as a stochastic process, where the randomness is driven by the environment, we write $\Psi(t,\xi(t))$ as a vector-valued random variable, which assumes the value $\psi(t,i)\in \bbH$ when $\xi(t)=i.$ 

To describe the time evolution of this system, note that before the first change of state of the environment, the $i$-th component evolves according to $H_i$ for each $i=1,..., K.$ To describe the joint evolution it is convenient to introduce the matrix (operator)-valued Hamiltonian:
\begin{equation}\label{ham1}
\bcH = \begin{pmatrix}
                      H_1& 0 & \dots &0\\
                      0  & H_2&\dots &0\\
                      \vdots& \vdots & \dots &\vdots\\
                      0 & 0 & \dots &H_K
                      \end{pmatrix}
\end{equation}
Similarly, if $\cA$ is a Hermitian matrix, describing an observable of the system, we denote by $\cbA,$ the operator valued, diagonal matrix, that has an $\cA$ along the diagonal.

Before a jump of state in the environment, each component evolves according to the Hamiltonian corresponding to that state, that is:
 
\begin{equation}\label{TE1}
\Psi(t,\xi(t)) = e^{-it\bcH/\hbar}\Psi(0,\xi(0)),\;\;\;t < T.
\end{equation}

The dynamics during the (random) time interval $[0,T)$ is, therefore, governed by
\begin{equation}\label{TE2}
i\hbar\frac{\partial}{\partial t}\Psi = \bcH\Psi,\;\;\; \Psi = \Psi(0,\xi(0)),\;\;\;t=0.
\end{equation}
What is important to realize, is that \eqref{TE1} means that on $[0,T)$ each component if 
$\Psi(t,X(t))$ is given by 
\begin{equation}\label{TE2}
\psi(t,\xi(t))= e^{-itH_i/\hbar}\psi(0,i) ;\;\;\;\mbox{if}\;\;\;\xi(t)=\xi(0)=i.
\end{equation}
And since the time evolution operator is continuous, $e^{-itH_i/\hbar}\to e^{-iTH_i/\hbar}$ as $t\to T.$
Also, since on $[0,T)$ we have $\xi(t)=\xi(0),$ we have $\xi(T-)=\lim_{t\uparrow T} \xi(t)=\xi(0).$ Therefore
\begin{equation}\label{TE3}
\lim_ {t \uparrow T}\psi(t,\xi(t)) = \psi(T,\xi(T-)) = \psi(T,x(0)).
\end{equation}
Using the notation introduced in \eqref{JT2}, the successive states of the environment are 
$$\xi(0), \xi(T_1), ...,\xi(T_n),..\;\;\;\mbox{during}\;\;\;[0,T_1),\;[T_1,T_2),...,[T_n,T_{n+1}),....$$
So, during the interval $[T_(n-1),T_n)$ the environment is in state $\xi(T_{n-1}).$ During this time interval, the quantum system evolves according to the Hamiltonian $H_{\xi(T_{n-1})}$ applied to $\psi(T_{n-1},\xi(T_{n-1})),$ that is:
 \begin{equation}\label{TE4}
\psi(t,\xi(t))= e^{-i(t-T_{n-1})tH_{\xi(T_{n-1})}/\hbar}\psi(T_{n-1},\xi(T_{n-1})) ;\;\;\;\mbox{for}\;\;\;T_{n-1}\leq t <T_n.
\end{equation}
At the end of the interval and {\it right} before the $n$-th jump of the environment, its state is $\xi(T_{n-1})$ and the state of the quantum system is 
$$\psi(T_n,\xi(T_{n-1}))= e^{-i(T_n-T_{n-1})tH_{\xi(T_{n-1})}/\hbar}\psi(T_{n-1},\xi(T_{n-1})).$$
 As when the environment makes a transition to $\xi(T_n),$ the quantum system suffers a shock (instantaneous perturbation) and becomes:
 \begin{equation}\label{TE5}
 \psi(T_n,\xi(T_{n})) = M(\xi(T_n),\xi(T_{n-1})\psi(T_n,\xi(T_{n-1})).
 \end{equation}.
 Here $\{M(j,i): 1 \leq, i,j \leq K\}$ denotes a collection of unitary operator acting on $\bbH.$ Contrary to the models in non-quantum random evolutions, the $M(j, i)'s$ have to be unitary to preserve the normalization of the state vectors. But as in the non-quantum case, they can be just a phase, or an operator acting on $\bbH.$ Below we will only consider the simplest of all cases, and suppose that $M(i,j)=V$ independently of the states $i$ and $j$ right before and after the transition. This operator will be augmented into an operator valued matrix $\cbV$ as time comes up.
 
 To continue, during $[T_n,T_{n+1})$ the quantum system will evolve according to the Hamiltonian $H_{\xi(T_n)}.$ If we use $\bbI_C$ for the indicator function of a set $C,$ we can write the random evolution (the time evolution along a random trajectory) of the quantum system as
 \begin{equation}\label{TE6}
 \Psi(t,\xi(t)) = \sum_{n=1}^\infty e^{-i(t-T_{n-1})tH_{\xi(T_{n-1})}/\hbar}\Psi(T_{n-1},\xi(T_{n-1}))\bbI_{[T_{n-1},T_n)}(t).
 \end{equation}
 
 It is convenient to have a shorthand notation for the time evolution operator applied to the initial state of the system. Let us write
 \begin{equation}\label{TE7}
 \Psi(t,\xi(t)) = \cbU(t,0)\Psi(0,\xi(0)).
 \end{equation}
 
 At this point, we mention that with some additional notations to denote the time-shift operations on the trajectories of the environment, \eqref{TE6} implicitly contains the path-wise first-time decomposition whose expected value, combined with the strong Markov property, yields \eqref{FTD1}.

\subsection{Time evolution in the Heisenberg picture}\label{HP}
When there is only one possible time evolution according to a Hamiltonian $H,$ in one single environment, and the system is prepared in an initial state $\psi(0),$ the expected value of an observable $A$ in state $\psi(t)$ at time $t$ is given by:
\begin{equation}\label{EVa}
\langle\psi(t),A\psi(t)\rangle = tr\big(A e^{-itH/\hbar}\rho(0) e^{itH/\hbar}\big);
\end{equation}
where $\rho(0)=\psi(0)\psi(0)^\dag=|\psi(0)\rangle\langle\psi(0)|.$ Using the notation $U^\sharp(t)(A)=U(t)AU^\dag(t)$  and the notation $H^\times(A)=[H,A]$ to describe its infinitesimal generator, i.e., $U^\sharp(t)=\exp(-itH^\times/\hbar),$ we can write \eqref{EVa} as: 

\begin{equation}\label{EVb}
\langle\psi(t),A\psi(t)\rangle = tr\big(A U^\sharp(t)\rho(0)).
\end{equation}
 
These notations allows to replicate verbatim the results in the previous section to describe the time evolution of the collection of density matrices associated with the different states of the environment. We use $\brho$ to denote the vector with components $\rho_i, i=1,...,K.$ This is a $K$-vector with $N-$dimensional density matrices as entries. When we need to consider of the components to be a density matrix-valued random variable, as above, we use $\brho_{\xi(t)}(t)=\brho(t,\xi(t))$ to describe the dependence in time and environment. 

Similarly, we have a time evolution operator $U^\sharp_i$ for every state of the environment. We gather them into an operator valued diagonal $K$-matrix $\bU_{\xi}^\sharp.$  For example, $\bU^\sharp(\brho_{\xi})(t)$ (respectively, $\cbA^\sharp(\brho_{\xi})(t)$) is an operator valued matrix, with elements $U^\sharp(\rho_i)(t)$ (respectively, $U^\sharp(\rho)_i(t)$) along the diagonal.

These notations allow us to express the result of the time evolution in a notationally friendlier (or more suggestive) form. For example, if the environment has gone through exactly $n$-shocks before time $t,$ at times $T_1<T_2...<T_n<t,$ and a state independent, instantaneous perturbation $V,$ acts when the shock occurs, then the density matrix at time $t$ is:

\begin{equation}\label{RE1}
\bU^\sharp_{\xi(T_n)}(t-T_n)\cbV^\sharp\bU^\sharp_{\xi(T_{n-1})}(T_n-T_{n-1})\cbV^\sharp....\cbV^\sharp\bU^\sharp_{\xi(0)}(T_1)(\brho(0)).
\end{equation}  

So, bringing in the indicator functions $\bbI_{\{N(t)=n\}},$ a reasonable notation for the time evolution in the random environment is

\begin{equation}\label{RE2}
\begin{aligned}
\brho(t,\xi(t)) &= \fU_{\xi(t)}^\sharp(t)(\brho(0))\\
&=\sum_{n=0}^\infty \bU^\sharp_{\xi(T_n)}(t-T_n)\bV^\sharp\bU^\sharp_{\xi(T_{n-1})}(T_n-T_{n-1})\bV^\sharp....\bA^\sharp\bU^\sharp_{\xi(0)}(T_1)(\brho(0))\bbI_{\{N(t)=n\}}.
\end{aligned}
\end{equation}  

That is what happens along each trajectory of the environment.  Next, to make predictions, we must average over all possible trajectories of the environment.

\section{Transition probabilities and expected values of observables}\label{EV}
Once we know how to obtain the state at time $t>0$ from the state $\psi_0$ at time $t=0,$ we can predict transition probabilities or compute expected values of an observable $\bA$ observables of interest. 

If there were no randomness in the environment, the predictor of the value of $\bA$ is given by: $\langle \psi(0)U(t)^\dag\bA U(t)\psi(0)\rangle = tr\big(\bA\bU^\sharp(t)\rho(0)\big).$ But as there are many possible time evolutions, we have to average over all of possible trajectories of the environment process, and the quantity that we need to compute is:
\begin{equation}\label{EV1}
\langle\bA(t)\rangle_{\psi} = E^{\xi(0)}[\tr\big(\bA\fU(t,0)\rho(0,\xi(0))\fU(t,0)^\dag\big)].
\end{equation}
We keep a possibly random, initial state $\xi(0)$ as initial state for the environment. If the reader so prefers, she might use a generic $i\in\{1,...,K\}.$ Using the notation of the last section, we have

\begin{equation}\label{EV2}
\langle\bA(t)\rangle_{\psi} = \tr\bigg(\bA E^{\xi(0)}{\xi(0)}[\cbU(t,0)^\sharp\big(\rho(0,\xi(0))\big)]\bigg) = tr\big(\bA E^{\xi(0)}[\brho(t,\xi(t))]\big).
\end{equation}

To follow the time evolution of the environment dependent density matrix, averaged over all the states of the environment, we must consider two cases, according to there being perturbation when the environment changes states or not.  First consider $E^{\xi(0)}[\brho(t,\xi(t))]$ when there is no perturbation when the environment changes states. As in Section 1.1, when the initial state of the environment is $\xi(0)=k,$ the first jump decomposition leads to:

\begin{equation}\label{EV3}
\begin{aligned}
&E^{k}[\brho(t,\xi(t))] = E^{k}\left[(\fU_{\xi(t)}^\sharp(t)\right](\brho(0))=e^{-t\lambda(\xi(0))}\bU^\sharp_{k}(\brho(0))\\
 &+ \lambda(k)\sum_{j\not=k)}^K Q(k,j)\int_0^t e^{-s\lambda(k}E^{j}[\bU^\sharp(t-s,\xi(t-s))]\bU^\sharp_{k}(\brho(0)).
\end{aligned}
\end{equation}

Differentiating with respect to $t,$ we obtain the differential equation:

\begin{equation}\label{VE4}
i\hbar\frac{\partial}{\partial t}\rho(t,k) = \bH_k^\times\rho(k) + i\hbar\sum_{j=1}^Nq(k,j)\rho(t,j),\;\;\;\rho(0,k)=\rho_k(0).
\end{equation}

To write this system in vector form, think of the transition rate as a matrix valued matrix $\bq,$ in which the element $q(i,j)$ is replaced by $q_(i,j)\bbI_N,$ because this matrix now is applied to a matrix valued vector $\brho,$ whose elements are $N-$dimensional matrices.  With all this, \eqref{VE4} in vector form looks like:
\begin{equation}\label{EV5}
i\hbar\frac{\partial}{\partial t}\brho = \bH^\times\rho + i\hbar\bq\brho,
\end{equation}
with prescribed initial condition $\rho(0,\xi(0)).$ 

To consider the case in which a perturbation occurs at the switching times, and furthermore, to keep the notations light, suppose that the perturbation is always the same, denoted by a unitary $\bV.$ In this case, the analogue of \eqref{EV3} looks like

\begin{equation}\label{EV6}
\begin{aligned}
&E^{\xi(0)}[\brho(t,\xi(t))] = E^{\xi(0)}\left[(\fU_{\xi(t)}^\sharp(t)\right](\brho(0)) = e^{-t\lambda(\xi(0))}\brho(t,\xi(0))\\
& + \lambda(\xi(0))\sum_{j\not=\xi(0)}^K Q(\xi(0),j)\int_0^t e^{-s\lambda(\xi(0)}E^{j}\left[\fU^\sharp(t-s,\xi(t-s))\right]\bV^\sharp\big(\brho_{i}\big)(s)ds.
\end{aligned}
\end{equation}

In differential form, this looks like \eqref{EV5}, namely:
\begin{equation}\label{EV7}
i\hbar\frac{\partial}{\partial t}\brho = \bH^\times\brho + i\hbar\bq\big(\bV^\sharp -\bbI\big))(\brho).
\end{equation}

It is of interest to know the Laplace transform of $E^{\xi(0)}\left[(\fU_{\xi(t)}^\sharp(t)\right].$ Let us compress the notations a bit and write $\langle\fU^\sharp\rangle(t)= E^{\xi(0)}\left[(\fU_{\xi(t)}^\sharp(t)\right].$ Introduce

\begin{equation}\label{LTEV1}
\langle\fU^\sharp\rangle(s) = \int_0^\infty e^{-st}\langle\fU^\sharp\rangle(t)ds.
\end{equation}
Keep in mind that $\brho(t,\xi(0))=\exp\big(-it\bH^{\times}/\hbar\big)\brho(0),$ and take the Laplace transform on both sides of \eqref{EV6}, drop the $\brho(0)$ from the identity, to obtain:

\begin{equation}\label{EV8}
\begin{aligned}
&\langle\fU^\sharp\rangle(s)= \big(s+\lambda + i\bH^\times/\hbar\big)^{-1}\\
& + \lambda(\xi(0))\sum_{j\not=\xi(0)}^K Q(\xi(0),j)\langle\fU^\sharp\rangle(s)\bV^\sharp \big(s+\lambda + i\bH^\times/\hbar\big)^{-1}.
\end{aligned}
\end{equation}

\section{Toy examples}
Below we consider two examples. The first one, which is a quantum mechanical version of Goldstein model, consists of a 2-dimensional quantum system in a two state Markovian environment. A notational variation of it, can be thought of as a modified 1-dimensional Dirac equation of a spinless particle. The second example consists of a 2-dimensional quantum system subject to random pulses in a one state environment, that is, after the pulse occurs, the quantum system evolves according to the same Hamiltonian, and at the shock it suffers an instantaneous perturbation consisting of an exchange of states. This is a toy version of the theme in \cite{CB}.

\subsection{2-dimensional quantum system in a two sate Markov environment}
Consider an environment with two states, and that the jumps between them occur with the same intensity $\lambda.$ Suppose furthermore, that the Hamiltonians describing the evolution in state 1 (respectively, state 2) are $H$ (respectively, $-H$). Think for example of $H=J\sigma_z.$ Of course, we are brushing the physical feasibility of the sudden transition from $H$ to $-H$ under the carpet.  This is simple analogue of the example considered in \cite{Gol}. See also \cite{Pin} for example. The objects introduced in Sections \ref{HP} and \ref{EV} are:

$$\langle\brho\rangle={\langle\rho_1\rangle\atopwithdelims()\langle\rho_2\rangle},\;\;\; \bH={H\,\;\;0\atopwithdelims()0\;-H},\;\;\;\bq={-\lambda\bbI_2\;\lambda\bbI_2\atopwithdelims()\lambda\bbI_2\;-\lambda\bbI_2}.$$
In this example, the version of \eqref{EV5} is a two dimensional system for a matrix valued vector:

\begin{equation}\label{Ex1.1}
i\hbar\frac{\partial}{\partial t}\langle\brho\rangle = \bH^\times\langle\brho\rangle + i\hbar\bq\langle\brho\rangle
\end{equation}
where $\langle\brho\rangle=E^{\xi(0)}[\brho(t.\xi(t))].$ 

The integration of this system proceeds as in Chapter 1 of \cite{Pin}. Keep in mind that when performing the following computations, the matrix products that appear below are ordinary products, despite the way that their entries act on the components of $\brho.$ Apply $i\hbar\frac{\partial}{\partial t},$ use the fact that
$$\bH^\times\bq+\bq\bH^\times = 2\lambda{-H^\times\;\;\;0\atopwithdelims() 0\;\;\;H^\times} = -2\lambda\bH^\times,\;\;\;\mbox{and}\;\;\;\bq^2= \lambda^2{\bbI_2\;\;-\bbI_2 \atopwithdelims()-\bbI_2\;\;\bbI_2},$$
and rearrange, invoke \eqref{Ex1.1}, cancel the common factor $(i\hbar)^2$ to obtain:

\begin{equation}\label{Ex1.2}
\frac{\partial^2}{\partial t^2}\langle\brho\rangle + 2\lambda\frac{\partial}{\partial t}\langle\brho\rangle = \big(\frac{1}{i\hbar}\bH^\times\big)^2\langle\brho\rangle.
\end{equation}

To leave the finite dimensional setup and relate to the Goldstein model, consider a 1 dimensional setup with classical Hamiltonian $H(x,p)=p$ in a two state environment. The quantized version of this Hamiltonian is $H=-i\hbar\partial/\partial x.$ This time the state space consists of 2-dimensional vectors with complex-valued functions as entries. The density matrices in this case are the usual $2\times2$-function valued matrices. With all this the analogue of \eqref{Ex1.1}, if we move the first term on the right to the left hand side, becomes:

\begin{equation}\label{Ex1.3}
i\hbar\begin{pmatrix}  
\frac{\partial}{\partial t}-\frac{\partial}{\partial x} & 0\\
0  & \frac{\partial}{\partial t}+\frac{\partial}{\partial x}
\end{pmatrix}\rho(t,x) =
i\hbar\lambda\begin{pmatrix} -1/2 & 1/2\\ 1/2 & -1/2\end{pmatrix}\rho(t,x).
\end{equation}

For a different variation on this theme, consider \cite{MO}. In the Goldstein model $H_1=v\partial/\partial x$ (the $i\hbar$ is absent), the analogue of \eqref{Ex1.2} is the telegraph equation. Observe that there $H_2=-H_1=-v\partial/\partial x$ can also be interpreted as generating the motion back in time. Setting $\bu=\exp(\lambda\hbar)\langle\brho\rangle,$ \eqref{Ex1.2} can be transformed into
a Klein-Gordon like equation. The interesting thing about \eqref{Ex1.2} is that the equations for the two components of $\bu$ (or $\brho$) are separated, but their solution are coupled through the initial condition for the time derivative at $t=0$ according to \eqref{Ex1.1}.

\subsection{Quantum system subject to Poisson distributed shocks}
For this example we consider a two dimensional system, evolving according to a Hamiltonian $H,$ and being subject to random pulses that occur according to a Poisson process of intensity $\lambda.$ Recall that $N(t)$ denotes the Poisson process that models the number of shocks. The Markovian structure of $N(t)$ is simple, since $N(t+s)-N(t)$ is independent of The behavior of the process up to time $t.$ If the a pulse occurs at time $T,$ and the system right before the pulse is $\psi(T-),$ the right after the pulse the state is $\bV\psi(T-),$ where the effect of the pulse is modeled by a unitary operator $\bV,$ so that the norm of $\psi$ is preserved. This extends to the Heisenberg picture in a natural way: If $\rho(T_)$ is the density matrix right before the shock, then right after, the 
 density matrix is $\bV^\sharp\rho(T-)=\bV\rho(T-)\bV^\dag.$

In this example, $\brho$ is just $\rho,$ and the average over environments is just the average over all possible paths of the Poisson process. The analogue of \eqref{EV6} and \eqref{EV7} are:

\begin{equation}\label{Ex2.1}
\begin{aligned}
E^{N(0)}[\rho(t,N(t))]& =  e^{-t\lambda}\rho(t,N(0))\\
& + \lambda \int_0^t e^{-s\lambda}E^{N(0)}[\big(\rho(t-s,N(t-s))\big)]\bV^{\sharp}\rho(t,N(0))ds.
\end{aligned}
\end{equation}
Here $\rho(t,N(0))=\exp(-stH^\times/\hbar)\rho(0).$ We use $N(0)$ as a reminder that at $t=0$ there has been no shocks upon the system, and the notation is convenient when using the Markov property. This equation can be used as a starting point for an iterative procedure to obtain the time dependence of $E^{N(0)}[\rho(t,N(t))].$ 

If, as above, we use $\langle\rho\rangle(t)=E^{N(0)}[\rho(t,N(t))],$ and differentiate with respect to $t,$ we obtain:

\begin{equation}\label{Ex2.2}
i\hbar\frac{\partial}{\partial t}\langle\rho\rangle = \bH^\times\langle\rho\rangle + i\hbar\lambda\big(\bV^{\sharp}-\bbI\big)(\langle\rho\rangle).
\end{equation}

Clearly, \eqref{Ex2.2} is not easy to integrate, except for particular choices of $\bV.$ But, we might follow \cite{CB} and consider the Laplace transform $\tilde{\rho(s)}$ of $\langle\rho(t)\rangle$ given by $\eqref{Ex2.1}.$ 
 Clearly, since the Laplace transform of a convolution is the product of the Laplace transforms of the factors, and all the operators in \eqref{Ex2.1} are linear, if we take the Laplace transform of both sides of \eqref{Ex2.1} we obtain:

\begin{equation}\label{Ex2.3}
\tilde{\langle\rho\rangle}(s) = \tilde{\langle\rho\rangle}_0(s+\lambda) - \big(s+iH^{\times}/\hbar + \lambda\big)^{-1}\lambda\bV^{\sharp}\tilde{\langle\rho\rangle}(s)
\end{equation}
Here we put
 
\begin{equation}\label{Ex2.4}
\tilde{\rho}_0(s+\lambda)=\int_0^\infty e^{-st}e^{-t\lambda}e^{-itH_0^\times}\rho_0 = \big(s+iH^{\times}/\hbar + \lambda\big)^{-1}\rho(0)=\big(s+iH^{\times}/\hbar+\lambda\big)^{-1}\rho(0).
\end{equation}
for the Laplace transform of the initial state evolved in time according to the free Hamiltonian $H_0.$ This approach is different but certainly related to the results obtained in \cite{CB}.

Under some simplifying assumptions, this example can be related to the theme of the previous example. Let us suppose that we choose a basis in which $\bH,$ $\bV$ and $\rho$ look like:

\begin{equation}\label{Ex2.5}
\bH = {E_1\;\; 0\atopwithdelims() 0\; E_2},\;\;\;\bV = {0\; 1\atopwithdelims() 1\; 0}\;\;\;\mbox{and}\;\\;\; \rho(0)={\rho_1(0)\;\;0 \atopwithdelims() 0\;\;\rho_2(0)}.
\end{equation}

As above, we apply $i\hbar\partial/\partial t$ to both side of \eqref{Ex2.2}, and bring it in again. If we use 
$$ \bH^{\times}\bV^{\sharp} = 0,\;\;\; (\bV^{\sharp}-\bbI)^2= 2(\bbI-\bV^{\sharp}),$$
then, after son arithmetical laboring, we obtain

\begin{equation}\label{Ex2.6}
\frac{\partial^2}{\partial t^2}{\rho_1\;\;0 \atopwithdelims() 0\;\;\rho_2} +2\lambda\frac{\partial}{\partial t}{\rho_1\;\;0 \atopwithdelims() 0\;\;\rho_2} = 4\lambda^2{\small{\rho_1-\rho_2}\;\; 0\atopwithdelims() 0 \;\; \small{\rho_2-\rho_1}}.
\end{equation}

Or, in components:
\begin{equation}\label{Ex2.7}
\begin{aligned}
\frac{\partial^2}{\partial t^2}\rho_1 +2\lambda\frac{\partial}{\partial t}\rho_1 = 4\lambda^2\big(\rho_1-\rho_2\big)\\
\frac{\partial^2}{\partial t^2}\rho_2 +2\lambda\frac{\partial}{\partial t}\rho_2 = 4\lambda^2\big(\rho_2-\rho_1\big)\\
\end{aligned}
\end{equation}
Now, adding and subtracting these two, we obtain a separated system of equations for $u_1=\rho_1+\rho_2$ and $u_2=\rho_1-\rho_2,$ which are coupled through their initial conditions (as is clear evaluating \eqref{Ex2.2} at $t=0$), which can be readily integrated.

\section{Closing remarks}
In the examples we considered two simple, extreme so to speak, cases. In the first, a two state environment that couples two different time evolutions, and in the other, one single state environment that causes a sudden change of state when a random pulse occurs. Clearly, the examples were chosen for their computational simplicity, and illustrate the possibilities of the class of models.

Among the possibilities for the occurrence of random pulses, one might as well suppose that the time between pulses and/or the effect of the pulse are experimental control parameters, which the experimenter might choose to obtain some specific response. This point of view might be worth exploring.

Another possible line of inquiry is related to the stochastic model for the random environment. As suggested by the models considered in \cite{Swi} or \cite{Da}, the nature of such models may have a wider range of applicability.

\textbf{Declaration of competing interests:} We have no competing interests to declare.

\end{document}